# Coarse Grained FLS-based Processor with Prognostic Malfunction Feature for UAM Drones using FPGA

Hossam O. Ahmed,

College of Engineering and Technology,

American University of the Middle East,

Kuwait.

*Abstract*— Many overall safety factors need to be considered in the next generation of Urban Air Mobility (UAM) systems and addressing these can become the anchor point for such technology to reach consent for worldwide application. On the other hand, fulfilling the safety requirements from an exponential increase of prolific UAM systems, is extremely complicated, and requires careful consideration of a variety of issues. One of the key goals of these Unmanned Air Systems (UAS) is the requirement to support the launch and control of hundreds of thousands of these advanced drones in the air simultaneously. Given the impracticalities of training the corresponding number of expert pilots, achieving this goal can only be realized through safe operation in either full-autonomous or semi-autonomous modes. According to many recent studies, the majority of flight accidents are concentrated on the last three stages of a flight trip, which include the Initial Approach, Final Approach, and Landing Phases of an airplane trip. Therefore, this paper proposes a novel decentralized processing system for enhancing the safety factors during the critical phases of Vertical and/or Short Take-Off and Landing (V/STOL) drones. This has been achieved by adopting several processing and control algorithms such as an Open Fuzzy Logic System (FLS) integrated with a Flight Rules Unit (FRU), FIR filters, and a novel Prognostic Malfunction processing unit. After applying several optimization techniques, this novel coarse-grained Autonomous Landing Guidance Assistance System (ALGAS3) processing architecture has been optimized to achieve a maximum computational processing performance of 70.82 Giga Operations per Second (GOPS). Also, the proposed ALGAS3 system shows an ultra-low dynamic thermal power dissipation (I/O and core) of 145.4 mW which is ideal for mobile avionic systems using INTEL 5CGXFC9D6F27C7 FPGA chip.

## I. INTRODUCTION

Sustainability for future generations also depend on finding innovative solutions to critical issues that we face nowadays, such as the rapidly increasing mobile population, and the concentration of migration into cities. Even though the current population has already added undeniable pressure to our existing infrastructure, the projections for the future show an even more devastating degradation of the quality of the most essential services such as transportation. Subsequently, research into innovative solutions for the expected issues in the transportation sector is a very active field of development.

One of the promising solutions proposed is via the Urban Air Mobility (UAM) initiatives [1-4]. One common thread in UAM proposals depends on creating a well-defined, and multi-level spatial network of virtual air routes, allowing the usage of autonomous taxi drones in under-utilized airspace domains, as a promising solution to the predicted surge of transportation demands of the future. On the other hand, there are enormous challenges to making such technology dependable. And importantly, these technology solutions must be even safer than conventional air-flighting systems, given an implicitly wider proliferation. Thus, reducing the accidents of UAM drones is a very high priority [5].

Breaking down the bigger challenges, into smaller solvable problems, can facilitate forward progress by taking into consideration the current key factors. Firstly, around 53.85% of the total current flight accidents occur during the last three stages of a flight trip; which include the Initial Approach, Final Approach, and Landing phases of an airplane [6, 7]. Secondly, we must consider the proliferation of UAM technology must necessarily be based on semi or fully autonomous flighting systems, given the impractical alternative of training expert pilots for such an enormous number of UAM drones per city. Thirdly, the planned UAM technology needs to consider that the routing of these trips will be dynamically assigned, and must be adaptable based on the variety of factors and scenarios that could occur spontaneously [8].

These factors could include natural or human-made ones, in addition to vehicle malfunctions, that might require immediately actionable emergency precautionary control procedures, referred herein as the "rules of flight". Considering these limitations leads us towards a better vision of how to make such UAM drones safer. Also, it should be considered that most of these taxi drones will depend on the Vertical and/or Short Take-Off and Landing (V/STOL) mechanism, which will increase the complexity of the electronic elements in the systems to guarantee the targeted safety factors [9]. Subsequently, we assume evolving research contributes to improving the safety measures for most of the flight stages through adoption of new high-tech approaches such as Machine Learning (ML), Deep Neural Networks (DNN), Neuromorphic Spiking Neural Networks (SNN), and other advanced control systems. However, we focused our efforts to enhance the safety precautions during the landing phases of a trip [10].

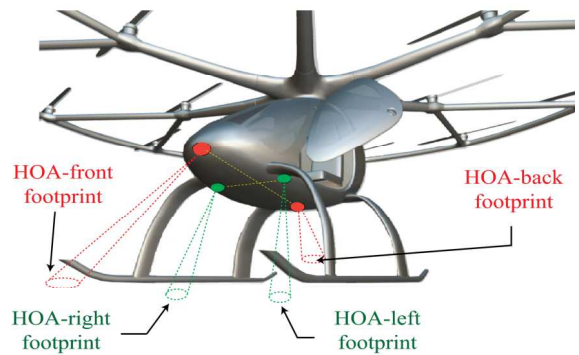

Fig. 1. The Graphical illustration of the proposed constellation of the two pairs of the differential HOA units.

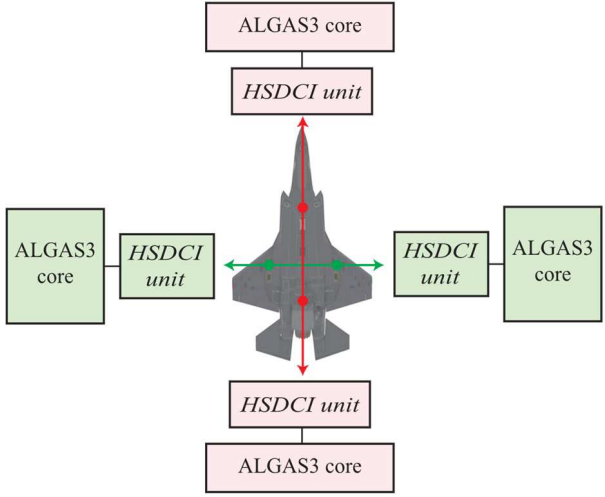

Fig. 2. The proposed distributed processing units of a complete ALGAS3 system.

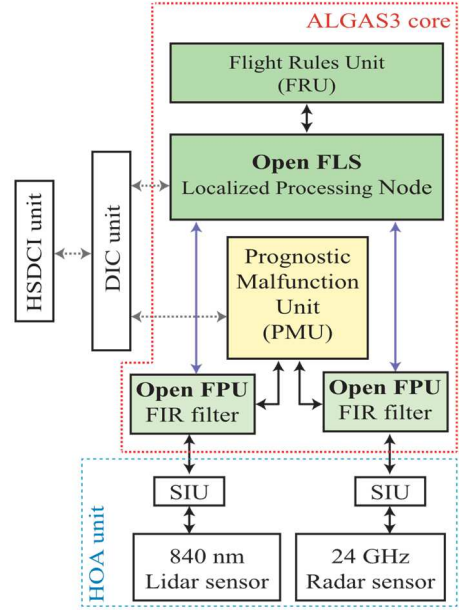

Fig. 3. The detailed structure of one spatial corner of the proposed ALGAS3 system.

The proposed vision can be achieved through open-architecture development of decentralized and collaborative processing cores to control the autonomous landing mechanism by receiving sensory data from the distributed Hybrid Obstacle Avoidance (HOA) sensory nodes allocated on the bottom side of a drone as illustrated in Fig 1. The proposed Autonomous Landing Guidance Assistance System (ALGAS3) processing architecture provides a real-time solution for avoiding landing accidents in different types of aircraft and automated drones. The proposed ALGAS3 processing unit depends on coarse-grained open-architecture Fuzzy Logic System (FLS) processing cores, and a configurable Flight Rules Unit (FRU). Also, other Digital Signal Processing (DSP) units are assumed to augment the proposed prognostic malfunction feature as will be explained in the upcoming sections. The proposed ALGAS3 processing architecture has been designed using the Very High-Speed Integrated Circuit Hardware Description Language (VHDL) in which approximate computing and a few other optimization techniques have been adopted to improve the dynamic power dissipation level and the computational speed performance. In section II of this paper, we review the related works, similar research, and contributions. In section III of this paper, we discuss the various architectural elements and techniques that have been used to boost the computational speed performance of the proposed ALGAS3 processing unit. In section IV of this paper, we discuss the results achieved. These were obtained after synthesis of the ALGAS3 architecture using Intel Quartus Prime tools.

The achieved computational speed performance of the ALGAS3 processing unit is about 69.89 Giga of Operations per Second (GOPS) using the INTEL Cyclone V 5CGXFC9D6F27C7 Field Programmable Gate Array (FPGA) chip. Also, we discussed the coding techniques that allowed the proposed ALGAS3 system to have an ultra-low dynamic thermal power dissipation (I/O and core) of 131.96 mW. And finally, in section V of this paper, we present our conclusions, and proposed future works.

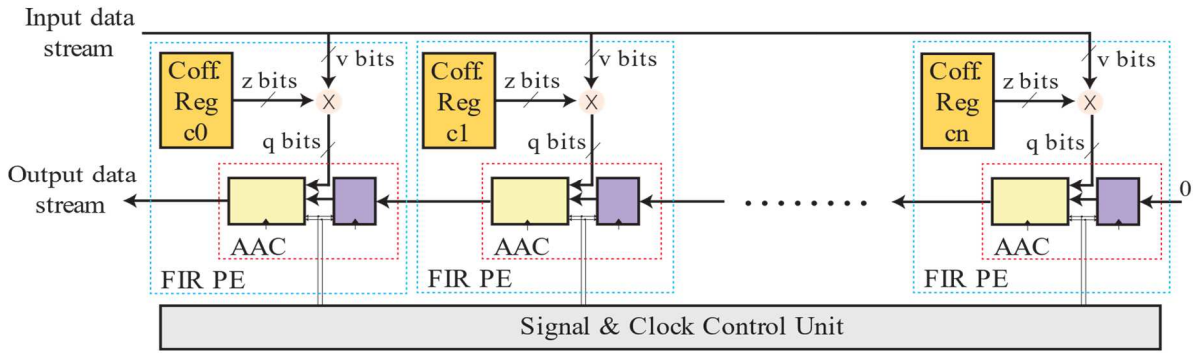

Fig. 4. The proposed block diagram of the coarse-grained architecture of the moving average FIR filter.

## II. RELATED WORK

The next generation of UAM needs more complex electronics systems to adhere with the flight safety regulations such as the Federal Aviation Administration (FAA) [11], Single European Sky ATM Research (SESAR), and the Next Generation (NextGen) air transportation Systems [12-14].

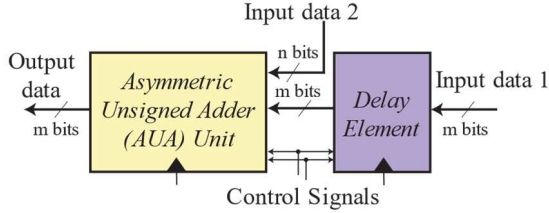

Fig. 5. The block diagram of the Adder and Accumulate (AAC) unit.

Many interesting contributions and research outcomes have been proposed to find innovative solutions to this issue by applying Artificial Intelligence such as Deep Neural Networks, Neuromorphic Systems, and Machine Learning; also, by using advanced control algorithms such as Fuzzy Logic Systems (FLS) [15-18]. Unlike Neural Networks, Fuzzy Logic Systems offer the advantage of applying pre-determined rules to imprecise data over variable conditions, without pre-training for every potential scenario. The usage of the FLS algorithm has proved its trustworthiness as a reliable solution to solve different issues related to UAM. The domains that the FLS could contribute substantially to this UAM field could cover autopilot dynamics, visual human tracking drone systems, etc. However, using FLS to reduce the self-landing and altitude failure dilemma could be the most requested demand for the UAM market in the soon future [19, 20].

Also, depending on a reconfigurable and powerful processing unit such as the FPGA could be considered as the anchor point for the development of such systems due its capability to adapt to the various changes that might be required based on the change of the flight terrain as an example. Importantly, the rules of flight can be reviewed and encoded in human-readable Fuzzy-Rule format, simplifying regional-authority governance, and rapid updates, in-sync with evolving compliance requirements. In general, the Fuzzy-Rules of any flight control system could be exemplified using fuzzy qualifiers (*Italics*) and logical semantics (in CAPS) as showing below:

- IF (Not Landing-Mode) AND IF (Region-Beacon-Signal is *Weak*) AND (Direction is *Away-From-Region-Beacon*) THEN (*further-reduce* Speed, signal Range-Limit-Error).

- IF (Landing-Mode) AND IF (Optical-Sensor is *Very Noisy*) AND (uWave-Sensor *is Very Noisy*) AND (UWB-Sensor *is Very Noisy*) THEN (Stop Landing-Mode, Enter Hover-Mode, signal Sensor-Error, enable Manual-Control).

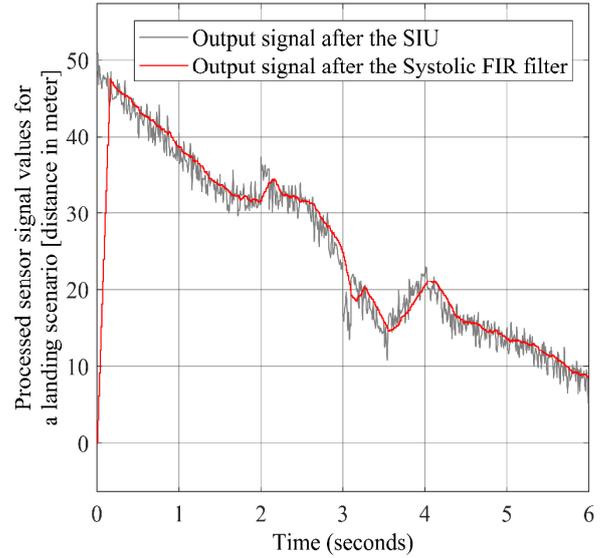

Fig. 6. The MATLAB validation of the Systolic FIR filter for a descending altitude scenario during a landing stage of a drone.

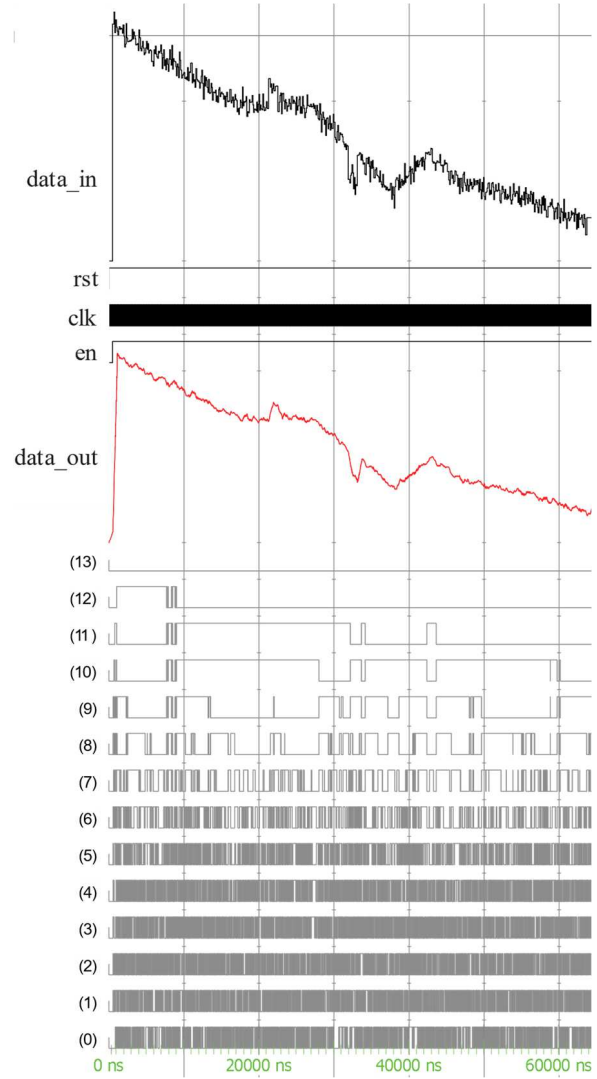

Fig. 7. The Questa Simulation validation of the Systolic FIR filter for a descending altitude scenario during a landing stage of a drone.

## III. THE PROPOSED ALGAS3 PROCESSING UNIT

The main contribution of this paper is to enhance the safety factors during the autonomous landing stages of the V/STOL taxi drones. This could be achieved by creating an autonomous system that could guarantee a stable landing procedure by the continuous distance measurements and data processing between the taxi drone and the landing area from four segregated HOA units and processing these aggregated data using the proposed ALGAS3 system. The proposed ALGAS3 system is a distributed and collaborative processing system that consists of four ALGAS3 cores as illustrated in Fig 2. Each two spatially opposite ALGAS3 cores are creating a differential processing pair to increase the safety factors by having a continuous doubled confirmation that the measured distance at these opposite sides of the taxi drone is matching to each other and is within the acceptable preset margin.

The communications and data exchange between these ALGAS3 cores are carried via the High-Speed Differential Comm Interface (HSDCI) unit. A detailed structure of one spatial corner of the proposed ALGAS3 system has been depicted in Fig. 3. Each ALGAS3 processing corner consists of four main subsections: The HOA unit, the ALGAS3 processing core, the HSDCI, and the Differential Inclination Control (DIC) unit. However, the focus of this paper is only on the ALGAS3 processing core. In general, the HOA unit is responsible for the data acquisition from the short-range 24GHz radar sensor and the 840nm lidar sensor via the Sensor Interface Unit (SIU). The selection of these two different sensors, to perform the same distance measuring task, is not only to increase the reliability of the measured distance but also to enhance the safety factor by applying the frequency spectrum separation concept. Hence, in case there is any spectrum attack in one of the bandwidths, the ALGAS3 system could give priority to the other sensor readings.

TABLE I. THE COMPUTATIONAL PERFORMANCE SUMMARY OF A SINGLE FIR MODULE AND A SINGLE PROGNOSTIC MALFUNCTION UNIT IN THIS PAPER.

|  | *Single FIR Module in this paper* | *Single Prognostic Malfunction Unit in this paper* |
|---|---|---|
| *FPGA device name* | INTEL 5CGXFC9D6F27C7 | |
| *Embedded FPGA's DSP resources usage* | None | |
| *Total dynamic core and I/O thermal power dissipation* | 16.45 mW | 11.44 mW |
| *External Memory usage or BRAM* | None | |
| *Max. freq.* | 253.1 MHz | 431.03 MHz |
| *No. of logic elements* | 155 ALMs | 46 ALMs |
| *Special feature per unit* | 15 TAPs | Fixed window size (frame of 16 samples) |

The ALGAS3 core consists of two systolic moving average Finite Impulse Response (FIR) filters to remove the noises from the sensory data of the two sensors in the HOA unit before being processed in the next stages. Each FIR filter has 15 TAPs. The proposed FIR filter consists of an Adder and Accumulate (AAC) unit and coefficient storage elements stage as shown in Fig. 4 and Fig. 5. The reason for this proposed coarse-grained architecture of the FIR filter is to be aligned with the other processing blocks in order to boost the computational speed of the next ALGAS3 version, as will be elaborated in the next proposed papers. The output signals of the two systolic FIR filters were fed to both the localized systolic FLS processing node and the Prognostic Malfunction Unit (PMU). The localized systolic FLS processing node is responsible for the data fusion task of the two sensory data to give output signals to the actuator control unit(s) for adjusting the altitude position of the drone itself during a landing mission.

TABLE II. THE COMPARISION BETWEEN THE PROPOSED ALGAS3 SYSTEM AND SIMILAR SIMILAR SYSTEMS.

|  | *The Proposed ALGAS3 in this paper* | *The Proposed ALGAS2 in [8]* | *The Proposed ALGAS1 in [9]* | *The Proposed FLS in [21]* |
|---|---|---|---|---|
| *FPGA device name* | INTEL 5CGXFC9D6F27C7 | | | |
| *FLS engines in the systems* | 4 Systolic basic cores + independent operation | 4 Systolic basic cores + independent operation | 5 Systolic based core | 3 systolic based core |
| *DSP units* | 2 FIR filters (15 TAPs) | None | | |
| *Prognostic malfunction feature* | One unit per ALGAS3 unit | None | | |
| *Total dynamic core and I/O thermal power dissipation* | 145.4 mW | 146.4 mW | 178.12 mW | 58.56 mW |
| *External Memory usage* | None | | | |
| *Max. freq.* | 276.63 MHz | 279.25 MHz | 266.03 MHz | 270.86 MHz |
| *FLS crisp inputs resolution* | Variable (11-bits, 10-bits) | | | |
| *No. of logic elements* | 4,544 ALMs | 3,488 ALMs | 4,304 ALMs | 2,578 ALMs |
| *Maximum Giga Operations Per second (GOPS)* | 70.82 GOPS | 21.22 GOPS | 25.273 GOPS | 14.36 GOPS |

The FRU is responsible for directing the Open FLS with the results of evaluating the IF-THEN-ELSE rule conditions. More details about the structure, simulations, and the analysis of this FLS unit could be found in [21-23]. The PMU is a novel processing element in the ALGAS3 core that could predict whether there is a significant drop in the sensory reading quality during the landing mission of the drone. Simply, it reads the sensory information of the two sensors and monitors the difference between these two signals over a certain time frame of 16 samples. Based on the outcomes of this process, the module could expect any urgent precautions that should be taken by the pilot directly. These precautions are indicating whether there is a failure of one of the sensors to provide a reasonable predicted data in comparison with the other sensor, or it could indicate that there is an intentional hacking (cyber-attack) at the bandwidth of one of the sensors.

## IV. THE RESULTS

To verify the proposed design, we divided the proposed system into two parts. The first part is related to the FLS unit that has been explicitly verified and analyzed in [20-22] as mentioned in section III. The second part is related to the newly proposed processing block that has been added to the ALGAS3 processing core such as the PMU and the systolic moving-average FIR filters as illustrated in Fig. 3. The validation of the systolic FIR filter design went through two stages. First, the design was validated using MATLAB, as shown in Fig. 6, by assuming a descending altitude scenario during a landing stage of a drone. The same signal from MATLAB has been fed as the input signals for simulating the VHDL version of the systolic FIR filter design using the Questa Simulation tool. As shown in Fig. 7, we achieved very satisfactory results in comparing with the desired assumptions. The validation of the PMU was completed using a different approach, in which we compared the simulation results of the VHDL version of this block, using the Questa Simulation tool, to the targeted mathematical model and conditions that have been assumed.

The achieved results were extremely matched to the expectations. To have a clear vision of the overall performance of the ALGAS3 system, we synthesized the entire system using the INTEL Quartus Prime tool. For better understanding the new introduced computational elements in the proposed ALGAS3 architecture, Table I is elaborating the computational performance for both the single FIR module and the Single PMU. Similar to all the other modules in the proposed ALGAS3 architecture, the FIR and the PMU are independent on the external memory usage or BRAM of the FPGA chip. Numerous logic improvements have been performed to decrease the amount of Adaptive Logic Module (ALM) resources required to implement the FIR and PMU on the FPGA device as depicted in Table I.

Consequently, both the FIR module and the PMU module have a low dynamic power consumption of 16.45 mW and 11.44 mW, respectively. As shown in Table II, the additional features that have been added to the ALGAS3 system cause it to utilize a total logic resource of 4,336 ALMs from the logic resources of the INTEL 5CGXFC9D6F27C7 FPGA chip. This increase by around 24.31% of the logic resources in comparison with the previous ALGAS2 system is reasonable since the computational speed performance escalated to 3.3x and 2.77x higher than the ALGAS2 system and ALGAS1 system respectively. Furthermore, the strategy of using different optimization techniques leads to achieving these promising results while having new features have been added to the system such as the prognostic malfunction feature and the Flight Rule Unit (FRU). The design of all the processing elements for the proposed ALGAS3 system has been accomplished at the gate level and RTL level using VHDL.

Moreover, the entire ALGAS3 system architecture has been optimized by depending on what we can call: "based-on-the-need" widths of all the buses, which has been briefly illustrated in the design of the systolic FIR filter in Fig. 4 and Fig. 5. This even helped us to enhance the total dynamic thermal power dissipation (I/O and core) to only 131.96 mW per the ALGAS3 system. Also, the focus on the dynamic power consumption in this paper is due to the fact of having meaningful information that will be used in the comparison with other similar ASIC-based systems in the future, since the static power consumption of the FPGA-based designs are representing the entire static power consumption per chip and not per the implemented system. Furthermore, Table II is showing more interesting features that have been taken into consideration for protecting the ALGAS3 system from any kind of fault injection attacks by eliminating the need to exchange the data with any type of external memory, and only the BRAM has been used for storing the coefficients.

## V. THE CONCLUSION

The paper introduced a new prognostic malfunction feature, and an Open FLS processing core to the ALGAS3 system that provides flexibility and transparent governance, which we assert are necessary for prolifically safer air travel in accordance with the rapidly evolving UAM guidelines from the FAA, the SESAR and NextGen safety measures during the autonomous landing of taxi drones. The ALGAS3 system presents a generic Cyber-Physical System (CPS-5) architecture that could enhance the safety of the taxi drones during landing operation. The current modifications to the ALGAS3 architecture will be the anchor point for achieving a more tangible processing performance in the futuristic version of this ALGAS3 architecture.

## VI. ACKNOWLEDGMENT

Many thanks to David Wyatt, IEEE CAS/DSASC member, for technical and support in the completion of this research project.